\documentclass{article}
\usepackage{spconf,amsmath,graphicx}
\usepackage{colortbl}
\usepackage{mathtools}
\usepackage{amsthm}
\usepackage{amsfonts,amscd,keyval}
\usepackage{bm}
\usepackage{multirow}
\usepackage{dsfont}
\usepackage{float}
\usepackage{xspace}
\usepackage{bbding}
\usepackage{graphics}
 \usepackage{enumitem}
 \usepackage{breqn}
\usepackage{booktabs}
\usepackage{xcolor}
\usepackage{hyperref}
\usepackage[square,sort,comma,numbers]{natbib}

\newcommand{\ours}{\textsc{DRAT}\xspace}

\DeclarePairedDelimiter{\nint}\lfloor\rfloor

\title{Dynamic Brain Transformer with Multi-level Attention for Functional Brain Network Analysis}
\name{{Xuan Kan$^{\dagger}$ \quad Antonio Aodong Chen Gu$^{\ddagger}$ \quad  Hejie Cui$^{\dagger}$ \quad  Ying Guo$^{\dagger}$\quad  Carl Yang$^{\dagger *}$\thanks{*Corresponding author: Carl Yang $<$j.carlyang@emory.edu$>$}}}
\address{$^{\dagger}$ Emory University, 
    $^{\ddagger}$ Georgia Institute of Technology}

\begin{document}
%
\maketitle
\begin{abstract}
Recent neuroimaging studies have highlighted the importance of network-centric brain analysis, particularly with functional magnetic resonance imaging. The emergence of Deep Neural Networks has fostered a substantial interest in predicting clinical outcomes and categorizing individuals based on brain networks. However, the conventional approach involving static brain network analysis offers limited potential in capturing the dynamism of brain function. Although recent studies have attempted to harness dynamic brain networks, their high dimensionality and complexity present substantial challenges. This paper proposes a novel methodology, Dynamic bRAin Transformer (DART), which combines static and dynamic brain networks for more effective and nuanced brain function analysis. Our model uses the static brain network as a baseline, integrating dynamic brain networks to enhance performance against traditional methods. We innovatively employ attention mechanisms, enhancing model explainability and exploiting the dynamic brain network's temporal variations. The proposed approach offers a robust solution to the low signal-to-noise ratio of blood-oxygen-level-dependent signals, a recurring issue in direct DNN modeling. It also provides valuable insights into which brain circuits or dynamic networks contribute more to final predictions. As such, DRAT shows a promising direction in neuroimaging studies, contributing to the comprehensive understanding of brain organization and the role of neural circuits.
\end{abstract}
\section{Introduction}
Network-centric analysis on brain imaging has gained substantial attention in neuroimaging studies recently, contributing profoundly to our understanding of brain organization in healthy individuals and those with brain disorders~\cite{brainnetworks}. Neuroscience research has consistently demonstrated that insights into neural circuits are pivotal for distinguishing brain function across diverse populations, with disruptions in these circuits often instigating and delineating brain disorders~\cite{insel2015brain}. Functional magnetic resonance imaging (fMRI) has emerged as a widely employed imaging modality for exploring brain function and organization~\cite{smith2012future}. Predicting clinical outcomes or categorizing individuals based on brain networks extracted from fMRI images with deep neural networks is a topic of significant interest in the neuroimaging community~\cite{BrainNetCNN_short, kan2022bnt, cui2022interpretable, zhu2022joint}.

\begin{figure}
     \centering
    \includegraphics[width=0.9\linewidth]{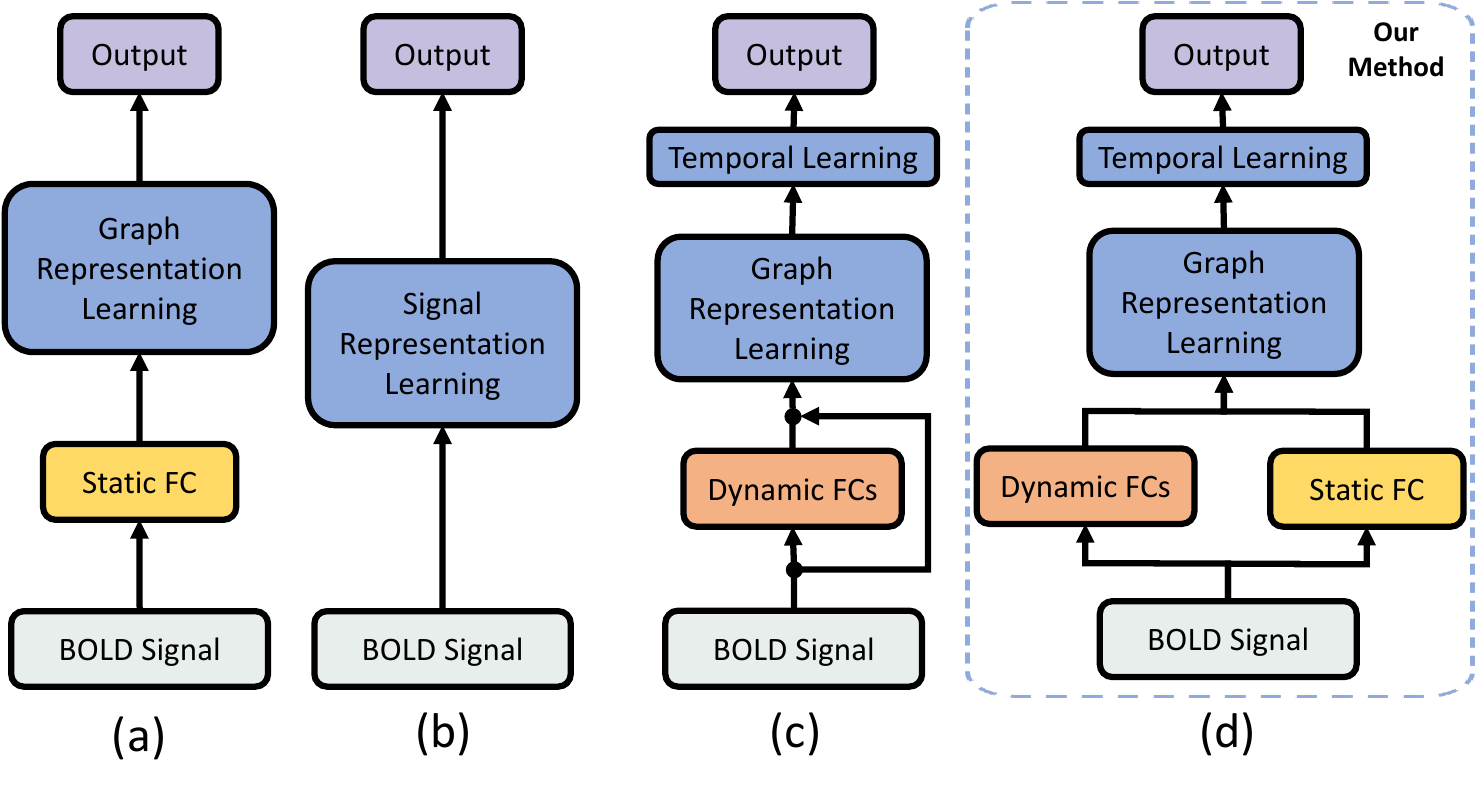}
    \vspace{-4ex}
    \caption{Four distinct schemas when employing Deep Learning for brain network analysis. From fMRI imaging, three types of input data can be acquired: (1) raw time-series data (BOLD signals), (2) static functional connectivity (FC), and (3) dynamic FCs, which capture temporal changes. Both kinds of FC are derived from the BOLD signal. Our method is the first attempt to combine Static FCs and dynamic FCs.}
    \label{fig:schema}
    \vspace{-10pt}
\end{figure}

Figure \ref{fig:schema} succinctly summarizes various schemas used for analyzing brain networks with neural networks. The classic approach to network analyses primarily relies on using individual fMRI data to construct functional brain networks~\cite{modellingfmri}. This established process involves selecting a brain atlas or regions of interest (ROI), extracting fMRI blood-oxygen-level-dependent (BOLD) signal series from each node or region, and computing pairwise connectivity measures. Once static brain networks are obtained, various neural network models can be applied for downstream analyses, as demonstrated in Figure \ref{fig:schema} (a). There have been attempts to model BOLD signals directly using deep neural networks (DNNs), as seen in Figure \ref{fig:schema} (b), but these have generally yielded unsatisfactory results due to the low signal-to-noise ratio of BOLD signals~\cite{kan2022fbnetgen, yu2023deep}. However, recent works have attempted to use dynamic brain networks to replace static ones for downstream analyses~\cite{LIU2023106521, kim2021learning}. These dynamic networks are created by segmenting BOLD signals into several overlapping or non-overlapping windows, each contributing to a unique connectivity matrix. This strategy shown in Figure \ref{fig:schema} (c) allows for exploring temporal variations and state transitions in functional connectivity over time, providing crucial insights into brain function. However, there is significant space for improvement due to the high dimensionality and complexity of dynamic brain networks. In response to these challenges, this paper proposes a novel methodology, Dynamic bRAin Transformer (\ours), depicted in Figure \ref{fig:schema} (d). \ours exploits static brain networks as a foundation measurement to integrate dynamic brain networks, thereby improving performance against benchmark methods. Additionally, we incorporate specific attention mechanisms to enhance model explainability, aiming to capitalize on dynamic brain networks' switching in neuroimaging studies.

\section{Method}

\begin{figure}
     \centering
    \includegraphics[width=0.8\linewidth]{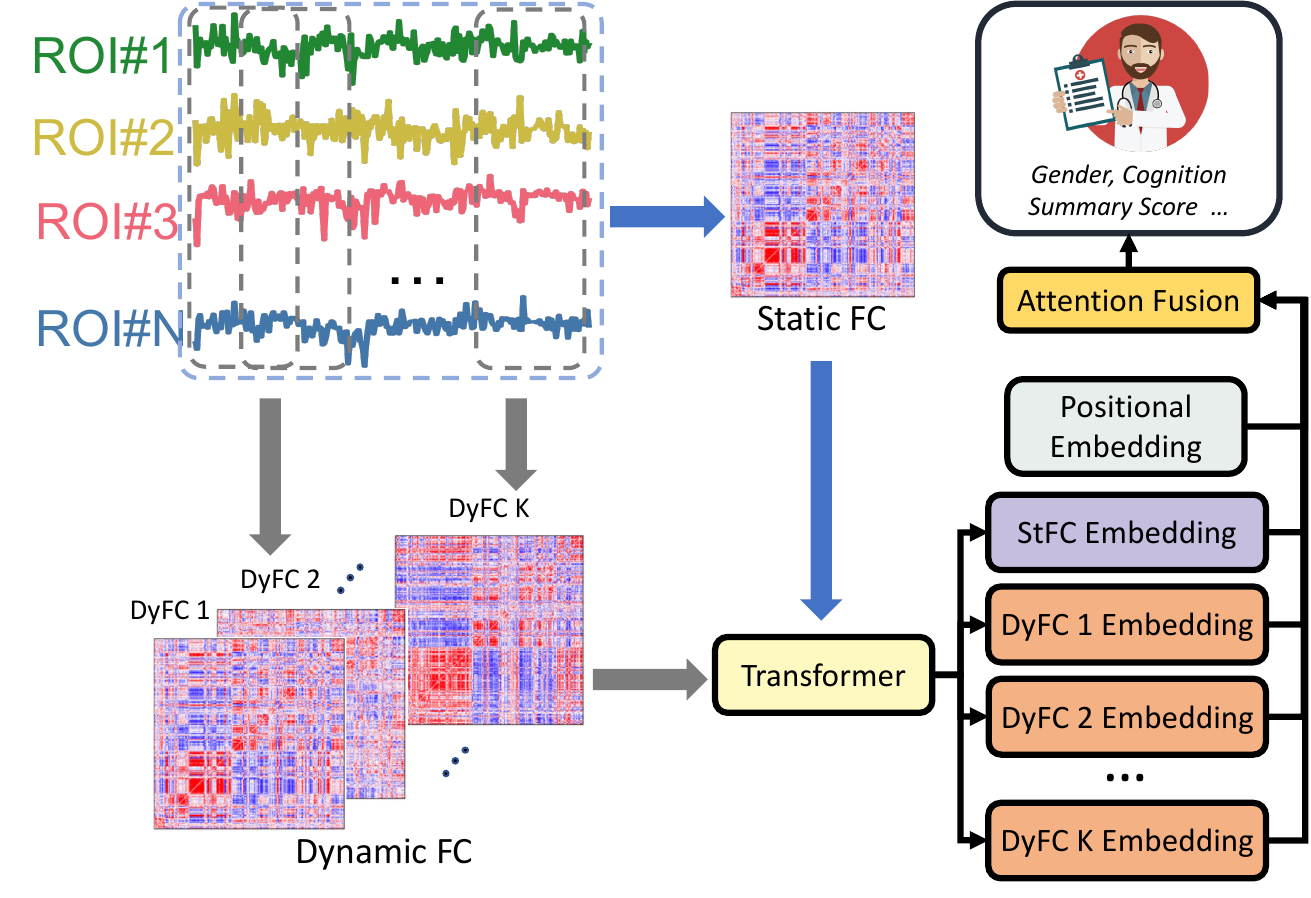}
    \vspace{-2ex}
    \caption{Diagram illustrating the comprehensive workflow of the proposed methodology, \ours.}
    \label{fig:flow}
    \vspace{-15pt}
\end{figure}

\begin{table}[htbp]
\vspace{-25pt}
\centering
\small
\caption{Performance comparison with baselines. The $\uparrow$ indicates a higher metric value is better, while $\downarrow$ is opposite.}
\label{tab:performance}
\resizebox{0.9\linewidth}{!}{
\begin{tabular}{cccccccc}
\toprule
\multirow{2.5}{*}{Type} & \multirow{2.5}{*}{Method} &\multicolumn{2}{c}{\bf PNC} & & \multicolumn{1}{c}{\bf ABCD}\\
\cmidrule(lr){3-4} \cmidrule(lr){6-6} 
& & {AUROC$\uparrow$} & {Accuracy$\uparrow$} & { } & {MSE$\downarrow$}\\
\midrule
\multirow{2}{*}{Dynamic}
&STAGIN & 63.5±4.0 & 54.2±1.4  & & 102.4±6.1 \\ 
&ST-GCN & 64.7±3.5&57.3±3.2  & & 89.2±11.2 \\ 
\midrule
\multirow{4}{*}{Static}
&BrainGNN & 62.4±3.5&59.4±2.3 && 80.8±4.7\\
& BrainGB      & 69.7±3.3 & 63.6±1.9   & & 78.1±4.3 \\
& BrainNetCNN  & 74.9±2.4 & 67.8±2.7   & &  77.1±4.5 \\
&BNT & 78.2±1.9 & 70.6±2.1  & & 60.2±1.5  \\ 
\midrule
\multirow{1}{*}{Dynamic \& Static} 
& \ours&  \textbf{80.7±3.1} & \textbf{72.5±2.3}  & &\textbf{ 58.3±3.5}  \\
\bottomrule
\end{tabular}
}
\end{table}
In this section, we elaborate on the design of \ours and its four main components as shown in Figure \ref{fig:flow}.
Specifically, the input $\bm X\in \mathbb{R}^{v \times T}$ denotes the BOLD time-series for regions of interest (ROIs) represents a sample (individual), $v$ is the number of ROIs, and $T$ is the length of time-series. We set $L$ as the window size, $S$ as the stride size. Given a sample $\bm X$, we can obtain $k$ dynamic brain networks, where $k=\nint{\frac{T-L+S}{S}}$.
For the classification task, the target output is the prediction label $\bm Y \in \mathbb{R}^{|\mathcal{C}|}$, where $\mathcal{C}$ is the class set of $Y$ and $|\mathcal{C}|$ is the number of classes. For the regression task,the target output is the prediction label $\bm Y \in \mathbb{R}$.

\noindent \textbf{Static FC Generation.} We begin by generating a static functional connectivity (FC) matrix, which provides a summary of the overall functional connections in the brain during the entire scan period. This static FC, $\bm A\in\mathbb{R}^{v\times v}$, represents the connectivity matrix between all pairs of ROIs for each individual. Specifically, we use the Pearson Correlation as a measure of statistical dependence between the time series of different ROIs. Each element of the static FC, $\bm A_{ij}$, is computed as $\text{Corr}(\bm X_i, \bm X_j)$, which denotes the correlation between the time-series of ROI $i$ and ROI $j$. This matrix captures the overall brain functional organization and serves as an anchor for the subsequent steps.

\noindent \textbf{Dynamic FC Generation.} To generate dynamic brain networks, we partition the BOLD signal into a series of overlapping or non-overlapping windows of length $T$, with a stride of size $S$. We calculate a dynamic functional connectivity matrix for the time window $t$, $\bm D^t \in \mathbb{R}^{v\times v}$. Finally, we can obtain $k$ dynamic functional connectivity matrix, where $k$ is the total number of dynamic networks given by $k=\nint{\frac{T-L+S}{S}}$. Each of these matrices represents a snapshot of brain connectivity at a specific time point. Similar to the static FC generation, we use Pearson Correlation for this computation. Each element in the dynamic connectivity matrix, $\bm D_{ij}^{t}$, is then calculated as $\text{Corr}(\bm X_{i}^{t}, \bm X_{j}^{t})$, where $\bm X_{i}^{t}$ and $\bm X_{j}^{t}$ are the BOLD signals for ROIs $i$ and $j$ at the time window $t$.

\noindent \textbf{Edge-level Attention and Transformer Projection.} After generating both static and dynamic FCs, we utilize the graph transformer proposed by~\cite{kan2022bnt} for processing these matrices. This transformer comprises a Multi-Head Self-Attention Module, which is adept at capturing complex dependencies between different nodes in the network, thus enabling a rich representation of the FCs. A learnable clustering readout function is applied to compress the matrix into a graph-level embedding. In the case of the static brain network, the hidden representation, $\bm h_A=\bm f_{\text{TF}}(A)$, is obtained. In contrast, for each dynamic brain network, the hidden representation is equipped with an attention layer $\bm \alpha$ and added by a positional embedding as delineated in \cite{vaswani2017attention}, resulting in $\bm h_D^t=\bm f_{\text{TF}}(\bm \alpha \circ \bm D^{t})+ \bm P^t$. Here, $\circ$ denotes the Hadamard product, $\bm \alpha \in \mathbb{R}^{v \times v}$ represents a learnable attention matrix (initialized at 1) shared across all dynamic networks, and $\bm P^t$ refers to the positional embedding at time window $t$. The attention mechanism based on $\bm \alpha$ allows the model to focus on the most informative connections in the brain networks.

\noindent \textbf{Temporal-level Attention and Fusion.} Given the hidden embedding of static FC  $\bm h_A$ and the sequence of dynamic FCs' hidden embedding $\bm h_D^t$, we utilize an attention mechanism to fuse these networks. The attention scores are computed based on the similarity between $\bm h_A$ and each $\bm h_D^t$. Specifically, the attention score $\bm \beta^{t}$ for each dynamic FC $\bm h_D^t$ is calculated as $\bm \beta^{t} = \text{softmax}(\text{sim}(\bm h_A, \bm h_D^t))$, where $\text{sim}(.)$ is a similarity function, such as dot product. Then, the fused FC $\bm F$ is generated as a weighted sum of dynamic FCs, i.e., $\bm F = \sum_{t=1}^{k} \bm \beta^{t} \cdot \bm h_D^t$. The final FC $\bm F$ is then fed into the multi-layer perception module for final prediction. The fusion mechanism thus enables the model to direct its focus towards dynamic FCs bearing higher similarity to the static FC, effectively leveraging the static FC's stable functional information to guide the dynamic FC fusion.
\section{Experiments}

\begin{figure*}
     \centering
    \includegraphics[width=0.80\linewidth]{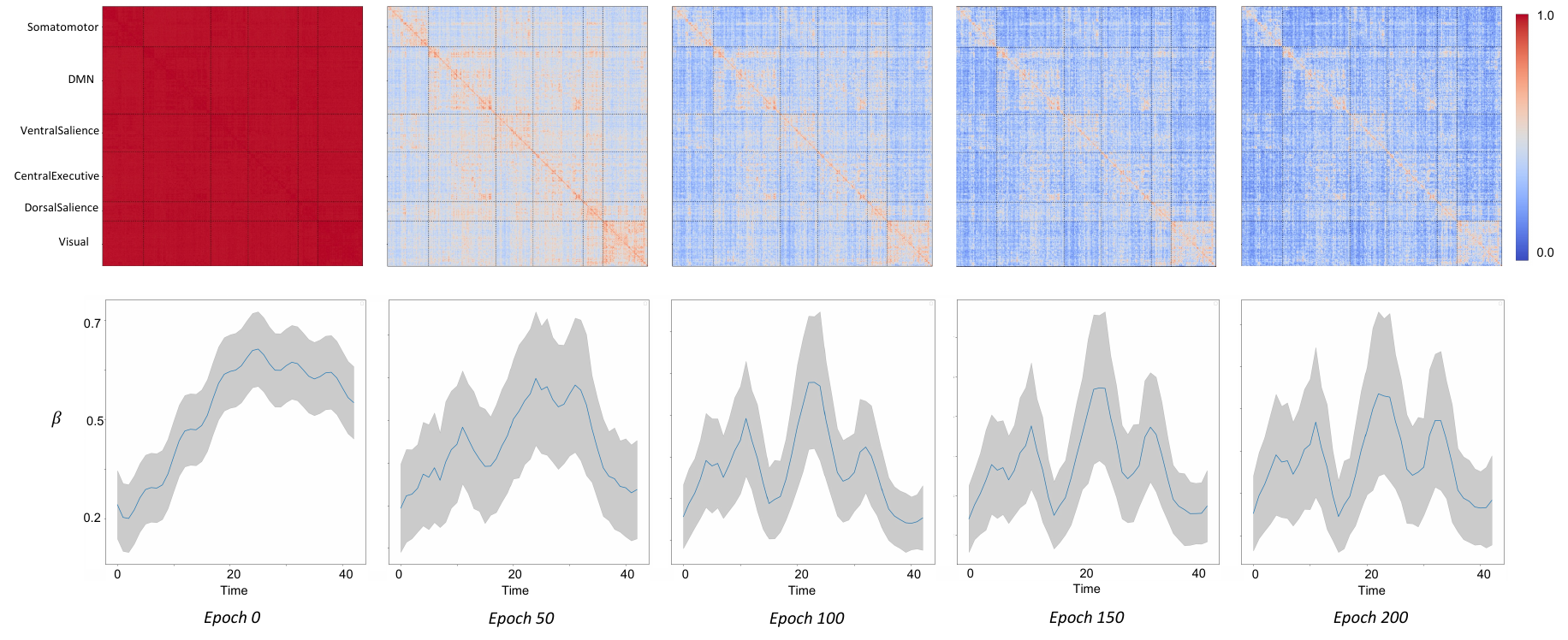}
    \vspace{-2ex}
    \caption{Evolution of two-level attention during the training on the ABCD dataset. The first row displays the progression of edge-level attention ($\bm \alpha$) across epochs, while the second row shows the changes in temporal-level attention ($\bm \beta$) across epochs.}
    \label{fig:attention}
    \vspace{-2ex}
\end{figure*}

\subsection{Experimental Settings}

\noindent \textbf{Dataset.} This study utilizes two public neuroimaging datasets. The first is the Adolescent Brain Cognitive Development Study (ABCD), one of the largest publicly available fMRI datasets with stringent access control~\cite{ABCD}. We employ fully anonymized brain networks based on the HCP 360 ROI atlas~\cite{GLASSER2013105_short} and define a task, the Cognition Summary Score Prediction, a regression problem focused on five cognitive sub-domains. Considering the variability in sequence lengths within the ABCD dataset, we included only those samples with a sequence length exceeding 1024, truncating them to form a unified length dataset, yielding 4613 samples for regression analysis.
The second dataset is the Philadelphia Neuroimaging Cohort (PNC), whose individuals aged 8–21 years provided by the Children's Hospital of Philadelphia~\cite{pnc_short}. After quality control, the dataset includes 503 subjects, each providing 120 timesteps of data from 264 nodes~\cite{power264_short}.

\noindent \textbf{Metric.} To evaluate performance in binary classification tasks, we employ two widely accepted metrics: Area Under the Receiver Operating Characteristic (AUROC) and accuracy. We set the classification threshold at 0.5 to determine the final class labels. In the case of regression tasks, we utilize the Mean Square Error (MSE) as a comprehensive measure of model performance. Please note, all the results presented in this study are the mean values derived from five independent runs, each initiated with a different random seed, to ensure the robustness and reproducibility of our findings.

\noindent \textbf{Implementation.}We configure the window size $L$ and stride size as 24 to ensure each window encapsulates a one-minute BOLD signal. The architecture of our Transformer is built according to the design described in \cite{kan2022bnt}, setting the number of transformer layers to 2, matching the hidden dimension for each transformer layer with the number of nodes $v$, and employing 4 heads. We divide our datasets such that 70\% is utilized for training, 10\% for validation, and the remainder for testing. We leverage the Adam optimizer throughout the training process with a learning rate and weight decay set at $10^{-4}$. Our batch size is 16, and all models undergo 200 training epochs. The epoch displaying optimal performance on the validation set is chosen for the final report.

\subsection{Performance and Analysis}

Our model's performance is benchmarked against several state-of-the-art methodologies in brain network analysis, and the result can be found in Table \ref{tab:performance}. We consider methods that leverage both static and dynamic brain networks for comparison. The baseline models include STAGIN~\cite{kim2021learning}, which constructs dynamic brain networks and fuses them using an attention mechanism without considering static brain networks; ST-GCN~\cite{gadgil2020spatio_short}, an improved version of GCN that takes into account not only the current graph but also the adjacency of prior and future graphs; BrainGNN~\cite{li2020braingnn_short} and BrainGB~\cite{cui2022braingb_short}, two Graph Neural Networks designed explicitly for static brain networks; BrainNetCNN~\cite{BrainNetCNN_short}, a convolutional neural network model designed for static brain networks; and finally BNT~\cite{kan2022bnt}, a graph transformer model also designed for static brain networks, which is the same transformer that we employ in our model to project brain networks into an embedding.

The comparative analysis with these methods demonstrates several vital insights. Models that rely exclusively on dynamic brain networks perform the poorest, underlining the critical role of global information provided by static brain networks in predictions. Static brain networks were found to encompass the most significant predictive signals, which illustrates our strategy of using the hidden representations of these static networks as anchor points (or query embeddings) to fuse dynamic brain representations. Furthermore, integrating dynamic brain networks is observed to enhance model performance because it exploits the fine-grained variations in brain states. As a result, our proposed methodology, denoted as \ours, consistently outperforms all baseline methods, exhibiting the highest performance across various metrics on two datasets with both regression and classification tasks.

\subsection{Attention Visualization and Analysis}

Due to the special attention design in \ours, the proposed method enables two-level attention-driven interpretability - edge-level attention ($\bm \alpha$), representing the significance of each edge, and temporal-level attention ($\bm \beta$), indicating the importance of each dynamic brain network. As evidenced in Figure \ref{fig:attention}, during the first 100 epochs, attention weights progressively evolve, stabilizing over the subsequent 100 epochs. The edge-level attention, initially uniform, progressively concentrates on the Visual and Default Mode sub-networks, aligning with the scores evaluated from heavy visual tasks like the Picture Vocabulary and Picture Sequence Memory Tests. Meanwhile, the temporal-level attention highlights dynamic brain networks recorded during the middle collection. Given the fact that the ABCD dataset is collected from children, it is plausible that resting-state brain activity is most prominent for children during the middle of the data collection process, corroborating our temporal attention insights.

\section{Conclusion}

Our proposed method, \ours, addresses the challenges presented by dynamic brain networks' complexity and high dimensionality, thereby advancing brain network analysis in neuroimaging. By leveraging static brain networks as a foundational measurement, we successfully integrate dynamic brain networks and improve performance compared to standard methods. The specific attention mechanisms incorporated within \ours further enhance model explainability. 




\label{sec:refs}

\bibliographystyle{IEEEbib}
\bibliography{refs}

\begin{thebibliography}{10}

\bibitem{brainnetworks}
Ed~Bullmore and Olaf Sporns,
\newblock ``{Complex brain networks: graph theoretical analysis of structural
  and functional systems},''
\newblock {\em Nature Reviews Neuroscience}, vol. 10, pp. 186--198, 2009.

\bibitem{insel2015brain}
Thomas~R Insel and Bruce~N Cuthbert,
\newblock ``Brain disorders? precisely,''
\newblock {\em Science}, vol. 348, pp. 499--500, 2015.

\bibitem{smith2012future}
Stephen~M Smith,
\newblock ``The future of fmri connectivity,''
\newblock {\em Neuroimage}, vol. 62, pp. 1257--1266, 2012.

\bibitem{BrainNetCNN_short}
Jeremy Kawahara et~al.,
\newblock ``{BrainNetCNN: Convolutional neural networks for brain networks;
  towards predicting neurodevelopment},''
\newblock {\em NeuroImage}, vol. 146, pp. 1038--1049, 2017.

\bibitem{kan2022bnt}
Xuan Kan, Wei Dai, Hejie Cui, Zilong Zhang, Ying Guo, and Carl Yang,
\newblock ``Brain network transformer,''
\newblock in {\em NeurIPS}, 2022.

\bibitem{cui2022interpretable}
Hejie Cui, Wei Dai, Yanqiao Zhu, Xiaoxiao Li, Lifang He, and Carl Yang,
\newblock ``Interpretable graph neural networks for connectome-based brain
  disorder analysis,''
\newblock in {\em MICCAI}, 2022.

\bibitem{zhu2022joint}
Yanqiao Zhu, Hejie Cui, Lifang He, Lichao Sun, and Carl Yang,
\newblock ``Joint embedding of structural and functional brain networks with
  graph neural networks for mental illness diagnosis,''
\newblock in {\em EMBC}, 2022.

\bibitem{modellingfmri}
Stephen~M. Smith, Karla~L. Miller, Gholamreza Salimi-Khorshidi, Matthew
  Webster, Christian~F. Beckmann, Thomas~E. Nichols, Joseph~D. Ramsey, and
  Mark~W. Woolrich,
\newblock ``{Network modelling methods for FMRI},''
\newblock {\em NeuroImage}, vol. 54, 2011.

\bibitem{kan2022fbnetgen}
Xuan Kan, Hejie Cui, Joshua Lukemire, Ying Guo, and Carl Yang,
\newblock ``Fbnetgen: Task-aware gnn-based fmri analysis via functional brain
  network generation,''
\newblock in {\em MIDL}. PMLR, 2022, pp. 618--637.

\bibitem{yu2023deep}
Yue Yu, Xuan Kan, Hejie Cui, Ran Xu, Yujia Zheng, Xiangchen Song, Yanqiao Zhu,
  Kun Zhang, Razieh Nabi, Ying Guo, Chao Zhang, and Carl Yang,
\newblock ``Deep dag learning of effective brain connectivity for fmri
  analysis,''
\newblock in {\em ISBI}, 2023.

\bibitem{LIU2023106521}
Lingwen Liu, Guangqi Wen, Peng Cao, Tianshun Hong, Jinzhu Yang, Xizhe Zhang,
  and Osmar~R. Zaiane,
\newblock ``Braintgl: A dynamic graph representation learning model for brain
  network analysis,''
\newblock {\em Computers in Biology and Medicine}, vol. 153, pp. 106521, 2023.

\bibitem{kim2021learning}
Byung-Hoon Kim, Jong~Chul Ye, and Jae-Jin Kim,
\newblock ``Learning dynamic graph representation of brain connectome with
  spatio-temporal attention,''
\newblock {\em NeurIPS}, vol. 34, pp. 4314--4327, 2021.

\bibitem{vaswani2017attention}
Ashish Vaswani, Noam Shazeer, Niki Parmar, Jakob Uszkoreit, Llion Jones,
  Aidan~N Gomez, {\L}ukasz Kaiser, and Illia Polosukhin,
\newblock ``Attention is all you need,''
\newblock {\em NeurIPS}, vol. 30, 2017.

\bibitem{ABCD}
B.J. Casey, Tariq Cannonier, and May I.~Conley et~al.,
\newblock ``The adolescent brain cognitive development (abcd) study: Imaging
  acquisition across 21 sites,''
\newblock {\em Developmental Cognitive Neuroscience}, pp. 43--54, 2018.

\bibitem{GLASSER2013105_short}
Matthew~F. Glasser et~al.,
\newblock ``The minimal preprocessing pipelines for the human connectome
  project,''
\newblock {\em NeuroImage}, vol. 80, pp. 105--124, 2013.

\bibitem{pnc_short}
Theodore~D. Satterthwaite et~al.,
\newblock ``{Neuroimaging of the Philadelphia Neurodevelopmental Cohort},''
\newblock {\em NeuroImage}, vol. 86, pp. 544--553, 2014.

\bibitem{power264_short}
Jonathan~D Power et~al.,
\newblock ``{Functional Network Organization of the Human Brain},''
\newblock {\em Neuron}, vol. 72, pp. 665--678, 2011.

\bibitem{gadgil2020spatio_short}
Soham Gadgil et~al.,
\newblock ``Spatio-temporal graph convolution for resting-state fmri
  analysis,''
\newblock in {\em MICCAI}. Springer, 2020, pp. 528--538.

\bibitem{li2020braingnn_short}
Xiaoxiao Li et~al.,
\newblock ``Braingnn: Interpretable brain graph neural network for fmri
  analysis,''
\newblock {\em Medical Image Analysis}, 2021.

\bibitem{cui2022braingb_short}
Hejie Cui et~al.,
\newblock ``{BrainGB: A Benchmark for Brain Network Analysis with Graph Neural
  Networks},''
\newblock {\em IEEE Transactions on Medical Imaging (TMI)}, 2022.

\end{thebibliography}

\end{document}